\documentclass[aps,prb,twocolumn,groupedaddress]{revtex4-1}
\usepackage[latin9]{inputenc}
\usepackage{graphicx}
\usepackage{epstopdf}

\makeatletter
 
 \@ifundefined{textcolor}{}
 {%
   \definecolor{BLACK}{gray}{0}
   \definecolor{WHITE}{gray}{1}
   \definecolor{RED}{rgb}{1,0,0}
   \definecolor{GREEN}{rgb}{0,1,0}
   \definecolor{BLUE}{rgb}{0,0,1}
   \definecolor{CYAN}{cmyk}{1,0,0,0}
   \definecolor{MAGENTA}{cmyk}{0,1,0,0}
   \definecolor{YELLOW}{cmyk}{0,0,1,0}
 }

\usepackage{dcolumn}
\usepackage{bm}

\makeatother

\begin{document}






\title{Theoretical Raman fingerprints of $\alpha$-, $\beta$-, and $\gamma$-graphyne}

\author{Valentin N. Popov}

\affiliation{Faculty of Physics, University of Sofia, BG-1164 Sofia, Bulgaria}

\author{Philippe Lambin}

\affiliation{D{\'e}partement de physique, Universit{\'e} de Namur, B-5000 Namur, Belgium}

\date{\today}
\begin{abstract}
The novel graphene allotropes $\alpha$-, $\beta$-, and $\gamma$-graphyne derive from graphene by insertion of acetylenic groups. The three graphynes are the only members of the graphyne family with the same hexagonal symmetry as graphene itself, which has as a consequence similarity in their electronic and vibrational properties. Here, we study the electronic band structure, phonon dispersion, and Raman spectra of these graphynes within an \textit{ab-initio}-based non-orthogonal tight-binding model. In particular, the predicted Raman spectra exhibit a few intense resonant Raman lines, which can be used for identification of the three graphynes by their Raman spectra for future applications in nanoelectronics.   

\end{abstract}
\maketitle


\section{Introduction}

The amazing physical properties of graphene\cite{novo04} have triggered the search for structures with superior performance among the graphene allotropes. One of their families, the graphynes, results from insertion of acetylenic groups $-$C$\equiv$C$-$ into carbon-carbon bonds of graphene, which can be done in a variety of ways.\cite{baug87,nari98} The graphynes are semiconductors with a finite gap\cite{nari98,colu03,kim12} or are, similarly to graphene, zero-gap semiconductors with linear electronic dispersion at the Fermi energy.\cite{malk12,kim12} It has been predicted that these structures share the unique mechanical,\cite{cran12} thermal,\cite{ouya12} and electrical\cite{chen13} properties of graphene.  Except for $\gamma$-graphyne, the vibrational properties of the graphynes have not been investigated yet.\cite{ouya12} Along with the theoretical study of the graphynes, there is some progress in the synthesis of fragments of graphene allotropes.\cite{malk12} Even though extended graphynes have not been synthesized yet, it is important to investigate theoretically their reponse to external perturbations for the needs of sample characterization for future nanoelectronics applications. A cheap and non-distructive characterization method is Raman spectroscopy, which has been applied with success to a number of all-carbon structures. The application of this method requires, however, the knowledge of the Raman bands of the graphynes.  

Here, we present a theoretical study of the Raman spectra of $\alpha$-, $\beta$-, and $\gamma$-graphyne within an \textit{ab-initio}-based non-orthogonal tight-binding (NTB) model. First, we calculate the electronic band structure and minimize the total energy to obtain the relaxed structure of the three graphynes. Then, the phonon dispersion and Raman intensity of the Raman-active phonons are derived using perturbation theory within the NTB model. Lastly, the simulated Raman spectra of the graphynes are discussed. The computational details are given in Sec. II. The obtained results are compared to available theoretical data in Sec. III. The paper ends up with conclusions (Sec. IV).  

\section{Computational details}

The NTB model\cite{popo04} utilizes Hamiltonian and overlap matrix elements, determined from \textit{ab-initio} data on carbon dimers.\cite{pore95} The total energy of the atomic structure is split into the sum of band structure and repulsive contributions. The latter is modeled by pair potentials with \textit{ab-initio}-derived parameters. The expression of the total energy is used for relaxation of the atomic structure, which is mandatory for the calculation of the phonon dispersion. The dynamical matrix explicitly accounts for the electronic response to the atomic displacements in first-order perturbation theory with electron-phonon matrix elements, calculated within the NTB model.\cite{popo10} Both the structure relaxation and phonon calculation require summations over the Brillouin zone of the graphynes. The convergence of the phonon frequencies to within 1 cm$^{-1}$ is reached on increasing the size of the Monkhorst-Pack mesh of points for the summations up to $40\times40$ ($\alpha$-graphyne), $15\times15$ ($\beta$-graphyne), and $20\times20$ ($\gamma$-graphyne). 

The Raman intensity of graphyne is calculated in fourth-order quantum-mechanical perturbation theory\cite{popo12} with  electron-photon and electron-phonon matrix elements, derived within the NTB model. The expression for the Raman intensity includes summations over the Brillouin zone. The integrated intensity of the Raman bands is found to converge to within a few percent by increasing the size of the Monkhorst-Pack mesh of points for the summations up to $200\times200$ for the three graphynes. The obtained Raman intensity depends on the laser photon energy. As a consequence, the Raman signal is resonantly enhanced for laser photon energies close to the optical transitions of the structure. Such a resonant behavior of the Raman scattering is normally observed in many all-carbon periodic materials, e.g., carbon nanotubes. In graphene, the scattering is resonant at all laser energies, which is manifested as a monotonous increase of the intensity with increasing laser photon energy. The resonant scattering makes it possible to observe Raman signal even from nano-size structures. 

\section{Results and Discussion}

\begin{figure}[t]
\includegraphics[width=80mm]{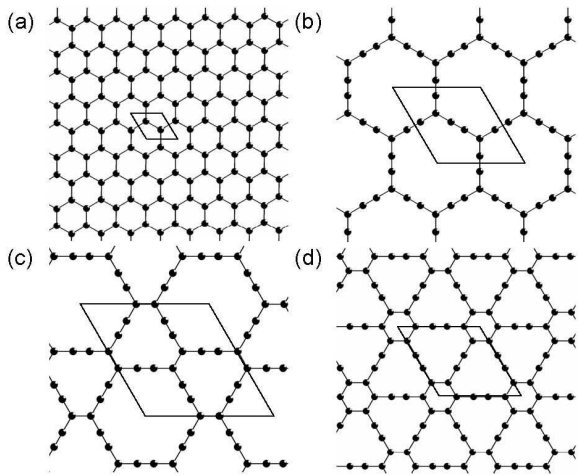} 
\caption{Schematic representation of the crystal structure of (a) graphene, (b) $\alpha$-graphyne, (c) $\beta$-graphyne, and (d) $\gamma$-graphyne. The atoms and bonds are denoted by solid circles and connecting lines, respectively; the rhombs are the unit cells of the structures.}
\end{figure}

\begin{table}[b]
\caption{\label{tab:table1}
Optimized unit cell parameter $a$ and bond lengths $r_{1}$, $r_{2}$, and $r_{3}$ (in $\text{\AA}$), graphyne unit cell area per atom relative to that of graphene, $RA$,  and relative excess of binding energy per atom (in eV) with respect to graphene, $\Delta E_{b}$, in comparison with available reported \textit{ab-initio} data. $r_{1}$ is the length of the bond between a triply-coordinated atom and its doubly-coordinated neighbors, $r_{2}$ is the length of the bond between two triply-coordinated atoms, and $r_{3}$ is the length of the bond between doubly-coordinated atoms (triple bond). For each graphyne, available reported \textit{ab-initio} data are provided.}
\begin{ruledtabular}
\begin{tabular}{rlllllll}
&$a$&$r_{1}$&$r_{2}$&$r_{3}$&$RA$&$\Delta E_{b}$&Ref.\\
\hline
$\alpha$&$6.992$&$1.402$&$-$&$1.232$&$2.02$&$0.96$&\\
&$6.997$&$1.4$&$-$&$1.244$&&&[\onlinecite{colu03}]\\
&$6.981$&$1.400$&$-$&$1.232$&&&[\onlinecite{kim12}]\\\
$\beta$&$9.507$&$1.415$&$1.398$&$1.225$&$1.72$&$0.87$&\\
&$9.46$&$1.43$&$1.34$&$1.2$&&&[\onlinecite{colu03}]\\
&$9.50$&$1.463$&$1.392$&$1.234$&&&[\onlinecite{kim12}]\\
$\gamma$&$6.894$&$1.430$&$1.407$&$1.220$&$1.31$&$0.63$&\\
&$6.86$&$1.42$&$1.40$&$1.22$&&&[\onlinecite{nari98}]\\
&$6.883$&$1.424$&$1.407$&$1.221$&&&[\onlinecite{kim12}]\\
\end{tabular}
\end{ruledtabular}
\end{table}

The $\alpha$-, $\beta$-, and $\gamma$-graphyne have the hexagonal symmetry of graphene itself with two-dimensional space group  $p6m$ (Fig. 1). The three graphynes can be derived from graphene by insertion of  an acetylenic group into every carbon bond, two-thirds of the carbon bonds, and one-third of the carbon bonds of graphene, respectively. Other, less symmetric graphynes can also be constructed by insertion of acetylenic groups. Nevertheless, the calculations for them can be performed in a similar way as for the considered graphynes.

The optimized unit cell parameter and carbon-carbon bond lengths are given in Table I along with the relative area per atom of the graphynes with respect to that of graphene, $RA$,  and the relative excess of binding energy per atom of the graphynes with respect to graphene, $\Delta E_{b}$. The lengths of the bonds between two triply-coordinated atoms and between a triply-coordinated atom and its  doubly-coordinated neighbors are close to that of graphene of $\approx1.42$ $\text{\AA}$ with a few exceptions (see Table I). The bond between doubly-coordinated atoms is $\approx1.2$ $\text{\AA}$, as expected for triple carbon bonds. The three graphynes have larger binding energy compared to graphene. Amongst the three graphynes, $\gamma$-graphyne has the smallest $\Delta E_{b}$ and the smallest $RA$, while these two quantities are the largest for $\alpha$-graphyne. For each graphyne, the optimized parameters agree well with the available \textit{ab-initio} ones in Table I. The deviation of the NTB structural parameters from the \textit{ab-initio} ones is comparable to the deviation between the two sets of \textit{ab-initio} parameters.  The reasons for the disagreement between the optimized parameters are unclear.

\begin{figure}[t]
\includegraphics[width=80mm]{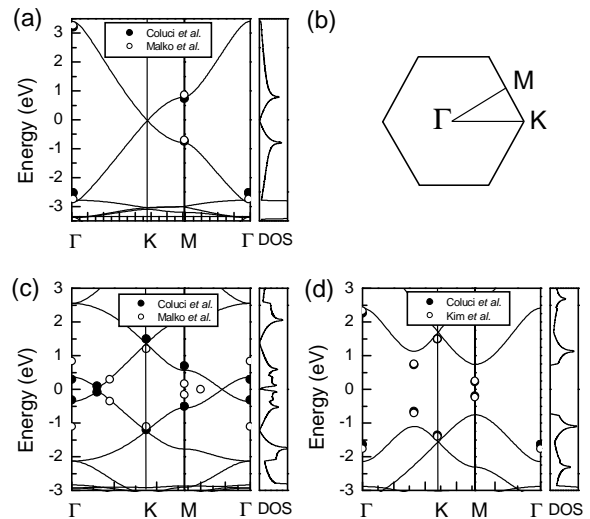} 
\caption{Electronic band structure and the density of states (DOS) of (a) $\alpha$-graphyne, (c) $\beta$-graphyne, and (d) $\gamma$-graphyne. Panel (b) shows the Brillouin zone of the three graphynes with high-symmetry points. The symbols are reported \textit{ab-initio} data.  In panel (c), the empty circle along the $\Gamma$M direction marks the predicted \textit{ab-initio} crossing of the conduction and valence bands and those along the $\Gamma$K direction mark the gap.\cite{malk12} The predictions of Ref.~\onlinecite{kim12} are similar to those of Ref.~\onlinecite{malk12} and are not shown on this panel.}
\end{figure}

Panels a), c), and d) of Fig. 2 show the calculated band structure of the three graphynes close to the Fermi energy (chosen as zero) and along three high-symmetry directions in the Brillouin zone (Fig. 2b). Graphene and the three graphyne structures sharing the same space group, the symmetry of the $k$ points in reciprocal space is the same as for graphene. According to symmetry, singly- and doubly-degenerate bands coexist at the K point of the first Brillouin zone for all these structures. Both $\alpha$-graphyne and graphene are honeycomb networks of carbon atoms and are characterized by similar band structures. Their Fermi level coincides with a doubly-degenerate band at the K point that is formed by the crossing of valence and conduction bands having linear dispersion when approaching the K point (Dirac cone). At the M point, the splitting between these two bands in $\alpha$-graphyne is $1.58$ eV, which agrees well with the reported value of $\approx1.6$ eV.\cite{colu03,malk12} In $\beta$- and $\gamma$-graphyne, the Fermi level does no longer coincide with a particular band at the K point. In $\beta$-graphene, there is a band crossing (at 0 energy) along $\Gamma$M that is allowed by the $C_{2v}$ symmetry of the Bloch vectors along this line. Additionally, a tiny gap of $0.16$ eV opens at a point along the $\Gamma$K direction, in agreement with the reported one of $\approx 0.15$ eV.\cite{colu03} By comparison, the corresponding bands in $\gamma$-graphyne are pushed away and do not cross. The calculated gap at the M point of $1.45$ eV corresponds well with the reported tight-binding result of $1.3$ eV (Ref.~\onlinecite{colu03}) but is about three times larger than the \textit{ab-initio} one of $0.39$ eV (Ref.~\onlinecite{colu03}) and $0.47$ (Ref.~\onlinecite{kim12}). This disagreement, partly due to density functional theory (DFT) underestimating the electronic band gap, affects the conditions for resonant Raman scattering for laser photon energy below the higher-energy gap.

The Fermi level of $\beta$-graphyne coincides with the energy of the band crossing along the $\Gamma$M direction discussed above. The Fermi surface is reduced to six points in the first Brillouin zone and, like graphene and $\alpha$-graphyne, $\beta$-graphyne is a zero-gap semiconductor.\cite{colu03,kim12,malk12} The shape of the Dirac cone is slightly more complex than in graphene and in $\alpha$-graphyne\cite{malk12} due to the lowering of symmetry, from threefold at the K point (trigonal wrapping) to mirror symmetry at the actual crossing point $k_F$ along the $\Gamma$M line. The exact location of $k_F$ depends on the details of the calculations. In a $\pi$ orthogonal tight-binding description,~\cite{kim12} $k_F$ depends on the relative values of the hopping parameters attributed to the three different C-C bonds that the $\beta$-graphyne structure has. In the present work, $k_F$ = $0.43 k_M$ where $k_M$ is the length of the $\Gamma$M segment. By comparison, H\"uckel-type calculations predict $k_F/k_M$ = 0.42,\cite{colu03} whereas \textit{ab-initio} calculations yield $k_F/k_M$ = 0.37,\cite{colu03} and 0.73.~\cite{kim12,malk12} The three cited \textit{ab-initio} calculations use DFT in the generalized gradient approximation and the Perdew-Burke-Ernzerhof exchange-correlation functional. The difference between the three calculations is in the program package used: SIESTA on the one hand\cite{colu03} and VASP on the other hand.\cite{kim12,malk12} In particular, the former package uses Troullier-Martins pseudopotentials, while the latter is based on the projector augmented waves method. The reasons for the disagreement between the available \textit{ab-initio} results should be elucidated by further sophisticated band structure studies. The reliability of the \textit{ab-initio} programs and the NTB model depends on the clarification of this point.

\begin{figure}[t]
\includegraphics[width=80mm]{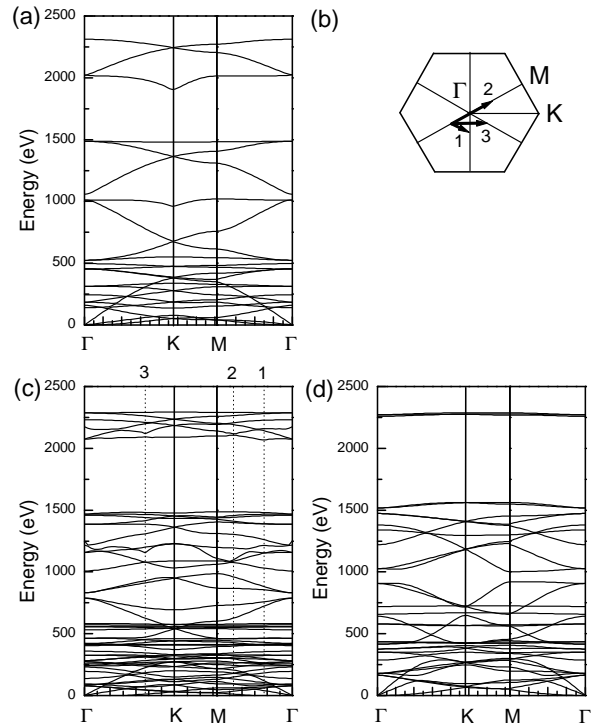} 
\caption{Phonon dispersion of (a) $\alpha$-graphyne, (c) $\beta$-graphyne, and (d) $\gamma$-graphyne. Panel (b) shows the Brillouin zone of the three graphynes with high-symmetry points. The solid arrows, labeled by $1$, $2$, and $3$, denote electron scattering paths between Dirac points at two points along the $\Gamma$M direction, which give rise to anomaly of the phonon dispersion of $\beta$-graphyne at three wavevectors. In $\alpha$-graphyne, the anomaly is observed at the K point due to electron scattering between Dirac points at two K points.}
\end{figure}

Next, we calculate the phonon dispersion of the graphynes. Before proceeding with the results, we note that the NTB model overestimates the phonon frequencies in comparison with the experimental ones. This has been demonstrated and discussed in the case of C$_{60}$.\cite{pore95a} The relative deviation of the NTB phonon frequencies with respect to the reported ones for C$_{60}$ (Ref.~\onlinecite{mene00}) increases with increasing frequency up to about $1000$ cm$^{-1}$ but remains almost constant in the high-frequency region above $1000$ cm$^{-1}$. Agreement of the calculated frequencies with the reported ones can be achieved by appropriate scaling of the former. We focus on the high-frequency region, which contains the intense Raman lines for the three graphynes, and derive a single scaling factor of $0.90$ with a standard deviation of $0.01$. The same factor has been found in graphene.\cite{popo10} We assume that this scaling factor is also applicable to the high-frequency phonons in the graphynes and scale all phonon frequencies by $0.9$. The accuracy of the scaled NTB phonon frequencies in the high-frequency region, deduced from the standard deviation of the scaling factor, is $\approx1\%$, which is an acceptable accuracy for the NTB model.

\begin{figure}[t]
\includegraphics[width=50mm]{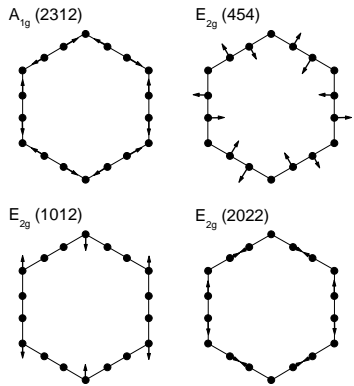} 
\caption{Atomic displacement of the Raman-active phonons of $\alpha$-graphyne, their symmetry species and frequency (in parenthesis, in cm$^{-1}$).}
\end{figure}

Panels a), c), and d) of Fig. 3 show the NTB phonon dispersion of the graphynes, scaled by $0.9$. As for graphene, there are three acoustic branches: a longitudinal and a transverse ones with linear dispersion and atomic displacements in the graphyne plane (in-plane phonons), and a transverse branch with  quadratic dispersion and displacements perpendicular to the graphyne plane (out-of-plane phonons). The latter branch determines the linear temperature dependence of the phonon heat capacity at low temperatures as in the case of graphene. The optical branches have in-plane and out-of-plane atomic displacements with bond-bending (displacements perpendicular to the bond) and bond-stretching (displacements along the bond) character. The bond-stretching phonons of the triple bonds form a strip of high-frequency branches. The calculated phonon dispersion of  $\gamma$-graphyne corresponds very well to the \textit{ab-initio} one,\cite{ouya12} which justifies the use of the scaling factor of $0.9$. 

According to their optical activity, the zone-center phonons can be Raman-active, ir-active, or inactive (silent). Here, we are concerned with the Raman-active phonons and their contribution to the Raman spectra. In the parent structure graphene, there is a single Raman-active phonon, the so-called $G$ mode, characterized by in-plane displacements of the carbon atoms. The Raman line, associated with this phonon is usually termed the $G$ band and is observed at $1582$ cm$^{-1}$. The atomic displacements of the Raman-active phonons of the graphynes are given in Fig. 4, 5, and 6, labeled by their symmetry in the hexagonal group $p6m$ and the calculated frequency (in parenthesis). There are altogether 4, 9, and 6 such phonons of the two symmetry species A$_{1g}$ and E$_{2g}$ in $\alpha$-, $\beta$-, and $\gamma$-graphyne, respectively. In all cases, the atomic displacements of the Raman-active phonons are in-plane ones. The phonons with frequencies below $\approx 1000$ cm$^{-1}$ have predominantly bond-bending character; the phonons with frequencies above $\approx 1000$ cm$^{-1}$ are predominantly bond-stretching. Among the latter, the phonons with frequencies above $\approx 2000$ cm$^{-1}$ are bond-stretching for the atoms involved in the triple bonds.

\begin{figure}[t]
\includegraphics[width=80mm]{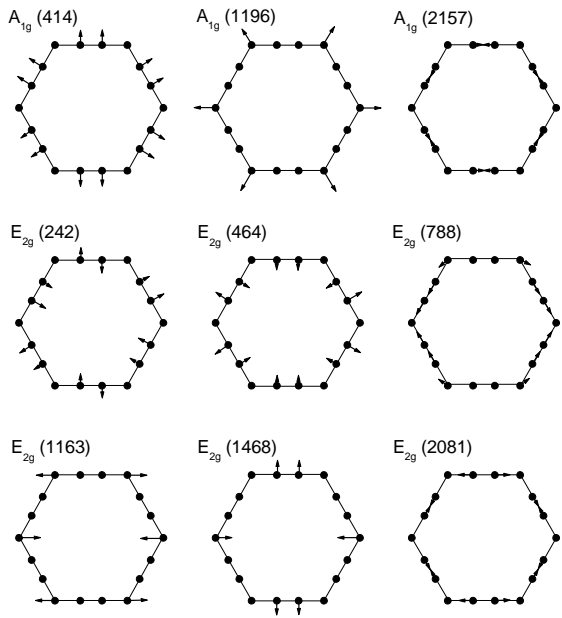} 
\caption{Atomic displacement of the Raman-active phonons of $\beta$-graphyne, their symmetry species and frequency (in parenthesis, in cm$^{-1}$).}
\end{figure}

The calculated Raman spectra of the graphynes for in-plane parallel scattering geometry at the commonly used laser photon energy $E_{L}=2.4$ eV are shown in Fig. 7. The spectra exhibit a few intense lines. In $\alpha$-graphyne, the two intense lines are due to phonons of E$_{2g}$ symmetry. The lower-frequency one at $1012$ cm$^{-1}$ arises from a G mode-like phonon. It is by $\approx50\%$ lower than the G mode of graphene, apparently due to softening of the restoring force by the replacement of $sp^{2}$ graphene bonds by single C$-$C ones. The higher-frequency line at $2022$ cm$^{-1}$ is due to a bond-stretching phonon with displacement of the atoms of the triple bonds. The other two Raman-active phonons give negligible contribution to the Raman spectra. In $\beta$-graphyne, the two intense lines come from phonons with A$_{1g}$ symmetry. The lower-frequency line at $1196$ cm$^{-1}$ is due to a phonon with breathing motion of the carbon hexagons. Contrary to $\alpha$-graphyne, the higher-frequency line is less intense than the lower-frequency one and arises from a phonon with bond-stretching movements of the triple-bond atoms. The Raman spectrum of $\gamma$-graphyne exhibits three intense lines, coming from the two phonons of A$_{1g}$ symmetry and one phonon of E$_{2g}$ symmetry. The Raman line at $1221$ cm$^{-1}$ is due to a phonon with breathing motion of the carbon hexagon and that at $2258$ cm$^{-1}$ is due to a phonon with displacements of the triple-bond carbon atoms. The Raman line at $1518$ cm$^{-1}$ is a G mode-like one but is softened to a lesser extent than in $\alpha$-graphyne, because of the replacement of only one $sp^{2}$ graphene bond by a single bond. The comparison of the three Raman spectra in Fig. 7, shows that the lowest-frequency lines are located in a narrow interval of Raman shifts around $1000$ cm$^{-1}$ and the highest-frequency have Raman shifts around $2000$ cm$^{-1}$. In order to distinguish between the samples of the three graphynes, Raman measurement in in-plane cross scattering geometry should be performed. In this geometry, only the E$_{2g}$ phonons in Fig. 7 should be observed (not shown).

\begin{figure}[t]
\includegraphics[width=80mm]{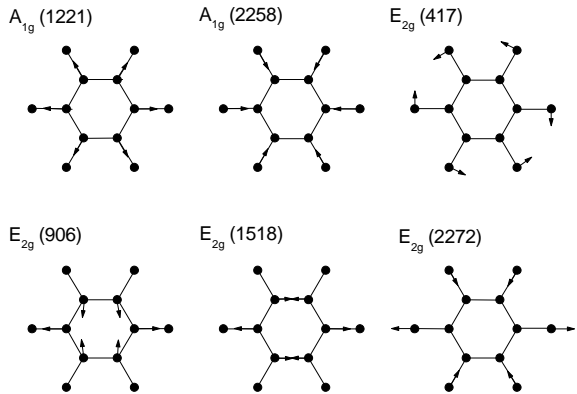} 
\caption{Atomic displacement of the Raman-active phonons of $\gamma$-graphyne, their symmetry species and frequency (in parenthesis, in cm$^{-1}$).}
\end{figure}

Finally, similarly to graphene,  defect-induced and two-phonon Raman bands are expected in the Raman spectra of $\alpha$- and $\beta$-graphyne.\cite{popo12} Such bands can originate from double resonant Raman scattering processes. In such a process, an incident photon is absorbed with creation of an electron-hole pair. The electron (or hole) is scattered by a phonon and a defect or by two phonons between the states of one (or two) Dirac cones before annihilation of the electron-hole pair with emission of a photon.\cite{popo12} The Raman shift of the most intense defect-induced and two-phonon bands can be deduced from the phonon branches with most prominent Kohn anomaly, which for both graphynes are along the $\Gamma$K direction. The length of the wavevector of the scattering phonons for double-resonant processes is given as $q=q_{DD}\pm E_{L}/v_{F}$, where $q_{DD}$ is the length of the wavevector, connecting two Dirac points and $v_{F}$ is the electron Fermi velocity at the Dirac point along the $\Gamma$K direction. The frequency of the scattering phonon is determined from the phonon dispersion, Figs. 3a and 3c. Thus, at  $E_L=2.4$ eV, two-phonon overtone bands should be observed at $\approx2030$ cm$^{-1}$ and $\approx4000$ cm$^{-1}$ for $\alpha$-graphyne, and at $\approx2450$ cm$^{-1}$ and $\approx4360$ cm$^{-1}$ for $\beta$-graphyne.  In presence of defects, defect-induced bands should be seen at about half of the Raman shift of the two-phonon bands. The thorough calculation of the defect-induced and the two-phonon bands is prohibitively time-consuming and is not done here.  This leaves open the question of the intensity of these bands.

\begin{figure}[t]
\includegraphics[width=80mm]{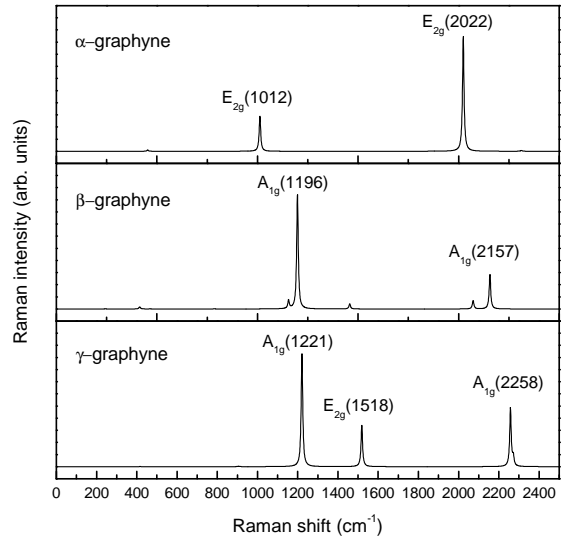} 
\caption{Raman spectra of $\alpha$-, $\beta$-, and  $\gamma$-graphyne at $E_L=2.4$ eV. The most intense lines of $\alpha$-graphyne, $\beta$-graphyne, and the $G$ band of graphene (not shown) have comparable intensity. The intensity of the lines of $\gamma$-graphyne strongly depends on $E_L$.}
\end{figure}

\section{Conclusions}
In conclusion, we have performed calculations of the electronic band structure, phonon dispersion, and Raman spectra of $\alpha$-, $\beta$-, and $\gamma$-graphyne. We have shown that among the many Raman-active phonons for each graphyne, only two or three of them give rise to intense Raman bands. The one-phonon Raman lines between $2000$ and $2300$ cm$^{-1}$ should be characteristic features of the graphynes. The predicted Raman bands can be used for  assignment of the lines of the Raman spectra of graphyne samples to particular graphynes.

\acknowledgments

V.N.P. acknowledges financial support from the Universit{\'e} de Namur (FUNDP), Namur, Belgium.

%


\end{document}